\documentclass[]{aastex631}

\submitjournal{PSJ}

\shorttitle{Geminids near PSP}
\shortauthors{Cukier \& Szalay}

\begin{document}

\title{Formation, Structure, and Detectability of the Geminids Meteoroid Stream}

\correspondingauthor{J. R. Szalay}
\email{jszalay@princeton.edu, jszalay@princeton.edu}

\author[0000-0002-8658-3811]{W. Z. Cukier}
\affiliation{Department of Astrophysical Sciences, Princeton University, Princeton, NJ, USA}

\author[0000-0003-2685-9801]{J. R. Szalay}
\affiliation{Department of Astrophysical Sciences, Princeton University, Princeton, NJ, USA}



\begin{abstract} 
The Geminids meteoroid stream produces one of the most intense meteor showers at Earth. It is an unusual stream in that its parent body is understood to be an asteroid, (3200) Phaethon, unlike most streams which are formed via ongoing cometary activity. Until recently, our primary understanding of this stream came from Earth-based measurements of the Geminids meteor shower. However, the Parker Solar Probe (PSP) spacecraft has transited near the core of the stream close to its perihelion and provides a new platform to better understand this unique stream. Here, we create a dynamical model of the Geminids meteoroid stream, calibrate its total density to Earth-based measurements, and compare this model to recent observations of the dust environment near the Sun by PSP. For the formation mechanisms considered, we find the core of the meteoroid stream predominantly lies interior to its parent body orbit and expect grains in the stream to be $\gtrsim$10 $\mu$m in radius. Data-model comparisons of the location of the stream relative to Phaethon's orbit are more consistent with a catastrophic formation scenario, in contrast to cometary formation. Finally, while PSP transits very near the core of the stream, the impact rate expected by Geminids meteoroids is orders of magnitude below the impact rates observed by PSP, and hence undetectable in-situ. We similarly expect the upcoming DESTINY+ mission to be unable to detect appreciable quantities of Geminids grains far from (3200) Phaethon.
\end{abstract}




\section{Introduction}
Meteoroid streams provide a discrete view into the dynamics and evolution of material in the zodiacal dust cloud. Much of our knowledge of meteoroid streams comes from Earth-based observations, where meteoroids ablate in the atmosphere and produce meteor showers observable via visual or radar-based systems. These detections rely on the Earth directly intersecting a substantial portion of the meteoroid stream and likely there are many streams in the solar system we do yet know of. 

{In this study, we focus on the Geminids meteoroid stream. It is an unusual steam as it is one of the strongest streams observed at Earth \citep[e.g.][]{jenniskens:94a} and is associated with asteroid (3200) Phaethon, instead of the typical cometary parent body of most streams. It has also been associated with asteroids 2005 UD and 1999 YC \citep[e.g.][]{jewitt:06a}, which may provide evidence of all 3 bodies being formed from a catastrophic breakup of an initial {comet} parent body.} The first well-documented observation of the Geminids occured in 1862 \citep{greg:1872a,king:26a} and has been studied considerably in the last half-century, both via observations \citep[e.g.][]{jenniskens:94a, arlt:06a, brown:08a, jewitt:10a, li:13a} and from a theoretical/modeling framework \citep[e.g.][]{fox:82a, fox:83a, ryabova:07a, ryabova:08a, ryabova:16a, ryabova:17a, ryabova:21a, williams:11a}. Many reviews of this shower have also been undertaken in the literature \citep[e.g.][]{zigo:09a, neslusan:15a, jakubik:15a, hajdukova:17a}. 

The Parker Solar Probe (PSP) spacecraft, launched in orbit about the Sun in 2018 \citep{fox:16a} is exploring the inner-most regions of our solar system's zodiacal dust cloud. {PSP's WISPR imager has made direct high-resolution visual images of the Geminids meteoroid stream very near to the Sun \citep{battams:20a} and discovered the core observable stream is offset radially outward from (3200) Phaethon's orbit. Additionally, PSP can directly observe its very local dust environment \citep{battams:22a}}. While it does not have a dedicated dust detector, it is able to detect the total impact rate of dust grains to the surface of the spacecraft on the FIELDS instrument \citep{bale:16a} via transient spacecraft potential changes due to the impact plasma from high-velocity impactors. The first few orbits of data were dominated by impacts from $\beta$-meteoroids \citep{szalay:20a}, fragments so small they are expelled from the solar system on unbound orbits due to radiation pressure. The {peak} fluxes of these grains were found to vary by $\sim$50\% over the first three orbits \citep{malaspina:20a}. Analysis of the directionality of impactors over the first few orbits also indicated the possibility of bound $\alpha$-meteoroids and retrograde impactors \citep{page:20a}. After PSP's initial orbit was successively lowered by flybys with Venus, comparison with a two-component model for the zodiacal cloud found its impact rates in subsequent orbits contained appreciable amounts of both $\alpha$-meteoroids and $\beta$-meteoroids \citep{szalay:21a}. A third source of impactors was identified that was not attributable to the nominal zodiacal cloud and was suggested to be either due to direct observations of meteoroid streams or their collisional byproducts in the form of a concentrated stream of $\beta$-meteoroids, a $\beta$-stream \citep{szalay:21a}. Broad estimates for the $\beta$-stream fluxes and analysis of the directionality of impactors \citep{pusack:21a} both favored a $\beta$-stream source over direct observations of meteoroid streams, but the exact origins of this third impactor source remain unresolved. 

While direct observations of meteoroid streams {have} been inferred by Helios data \citep{kruger:20a}, the densities required for such an observation and for PSP \citep{szalay:21a} are on the order of 10-100 km$^{-3}$. Here, we create a dynamical model of the Geminids with {three} simplified formation mechanisms to test if the Geminids meteoroid stream could be directly detectable by PSP and explain the unresolved third impactor source. We also investigate the structure of the stream and how it is dependent on the specific formation mechanism, {specifically in comparison to PSP's remote observations showing the stream lies outside its parent body orbit \citep{battams:22a}}. In Section~\ref{model} we describe the dynamical model. We compare the model to Earth-based and PSP observations {in} Section~\ref{comparison} and conclude with a discussion on implications and implications of model assumptions in Section~\ref{conclusions}.

\section{Simulation Methodology} \label{model}
The mass of the Geminids stream is estimated to be on the order of or larger than the parent body (3200) Phaethon \citep{blaauw:17a,ryabova:17a}, which suggests it was formed in a possibly catastrophic event that shed a large amount of mass in a relatively short period of time approximately 2000 years ago \citep{fox:83a,ryabova:99a}. A rapid cometary hypothesis has been shown to reproduce aspects of the Earth-based observations of the Geminids \citep{ryabova:21a}. While there is cometary-like activity currently occurring at (3200) Phaethon, the mass loss from this activity is orders of magnitude too low to sustain the Geminids \citep{jewitt:10a, jewitt:13a, jewitt:18a, tabeshian:19a}. Additionally, {(3200) Phaethon} will also experience an intense impact environment from co-orbiting dust in the zodiacal cloud near each perihelion, which ejects {sub-micron to mm-sized} material from the surface as well, but this mechanism is only estimated to be produce {$10^5-10^6$ kg of ejecta over 2000 years \citep{szalay:19a}, orders of magnitude too little material to appreciably contribute to the stream mass of $10^{13}-10^{15}$ kg \citep{blaauw:17a,ryabova:17a}}. Electrostatic dust ejection has been proposed as a  mechanism for the observed activity of (3200) Phaethon and possibly also as a source for the entirety of the Geminids stream \citep{kimura:22a}. Here, we simulate the formation of the stream as a catastrophic event occuring 2000 years ago and do not consider the effects of an ongoing source from (3200) Phaethon.

We consider three simplified creation scenarios for the Geminids stream: a) all the particles were released at (3200) Phaethon perihelion at velocity of 0 relative to the parent body as a basic model for a basis of comparison, b) all the particles were released at (3200) Phaethon perihelion with a velocity {dispersion on the order of 1 km/s} relative to the parent body to simulate a violent creation event and c) an elementary cometary model where the particles are released throughout a single orbit of (3200) Phaethon at a rate inversely proportional to the distance from (3200) Phaethon and the Sun. 


\subsection{Basic Model}
{We simulated 10,000 particles for each of} the {B}asic {Model} and {the V}iolent {C}reation {M}odel. These particles were released from perihelion at the same location and velocity as (3200) Phaethon 2000 years ago. Given the particles were released at perihelion, only $\beta$-values low enough that particle escape would not happen needed to be simulated.  Following \citet[Eq. 23,][]{burns:79a}, grains released from perihelion with $\beta \geq (1-e)/2$ are unbound and travel in hyperbolic orbits. The maximum $\beta$-value that need be considered for particles released from (3200) Phaethon at perihelion is $\beta = 0.052$, {which for the ``asteroidal'' model from \citet{wilck:96a} {with} particle density of} $\rho = 3205$ {kg/m}$^3${, corresponds to particles with a mass of }$2.4 \times 10^{-9}$ {g and radius of 12} $\mu${m}.  For the {B}asic {M}odel, the 10,000 particles were chosen to be evenly distributed in $\beta$-space from $\beta=0$ to $\beta=0.052$. 

The simulation was run from (3200) Phaethon perihelion on Jan 2, 19 A.D. for 2000 years using  \texttt{Rebound} N-body code \citep{rebound}. The simulations were integrated using IAS15, a 15th order Gauss-Radau integrator \citep{reboundias15}.  Ephemeris of (3200) Phaethon from 1600 A.D. was taken from JPL HORIZONS and back-integrated to the start date.  In this scenario, particles were released at perihelion with the same state vector as (3200) Phaethon.  

For each model, \texttt{Reboundx} \citep{2020MNRAS.491.2885T} was utilized to calculate the forces of radiation pressure and Poynting-Robertson drag.  Gravitational forces from the Sun, 4 inner planets, and Jupiter were considered in the simulation, data for which was taken from NAIF \texttt{SPICE} kernels \citep{FolknerPlanetaryLunarEphemerides}.  {We used only 5 instead of a full 8 planets to save on computational time. A test run of the basic model with all 8 planets caused only minor changes--a slight antisunward shift in the stream at perihelion and a change in the predicted stream mass that was significantly smaller than other uncertainties.} Dust particles were treated as test-particles that do not interact with each other.  For some configurations of parameter space, some particles would either get ejected from the solar system or become unphysically close to the Sun {where they would breakup via collisions or sublimation \citep[e.g.][]{mann:04a}. Here, we do not include the effects of sublimation/collisions and instead remove grains with very small semi-major axes from our simulation}.    {Particles with a semi-major axis of 0.2 au or below were removed from the simulation under the assumption that their interactions would be dominated by zodiacal collisions, sublimation and/or sputtering very near to the Sun. Additionally, particles that during any time step were less than $0.01$ au or greater than $10$ au from the Sun were removed from the simulation due to numerical concerns.} For each of the first 1998 years of the simulation the semi-major axis, eccentricity, inclination, argument of perihelion, and longitude of ascending node were stored for each particle.

During the last two years of the simulation, the position vector for each particle was stored at 2000 instances equally spaced in time.  Each of these snapshots was then treated as a separate particle for the remainder of the analysis.  Particles were then weighted {according to a mass power law as follows:}  $\beta$-mass relations were taken from \citet[Fig. 2,][]{wilck:96a}, treating the grain compositions as typically asteroidal, and extrapolated using a power law fit to the descending half of the curve (from $\sim 10^{-13}$ g to $\sim 10^{-{8}}$ g).  

The differential mass distribution for the Geminids stream is assumed to follow a power-law,
\begin{equation}
    f(m) \propto m^{-s}
\end{equation}
where $s$ is taken to be $1.68 {\pm 0.04}$ \citep{blaauw:17a}.  Integrating this differential mass distribution from a minimum mass{,} $m_0${,} to $m$ the cumulative mass distribution for the stream can be modeled as:
\begin{equation}\label{eqn:cume_mass}
    F(m) = \frac{m^{1-s} - m_0^{1-s}}{m_1^{1-s}-m_0^{1-s}}
\end{equation}
where $m_0$ was $10^{-9}$ g, below which particles released at perihelion would be ejected from the solar system and the maximum mass, $m_1$ was chosen to be $10$ g.

Particles were then weighted by first converting the {10,000} $\beta${-values} into mass {values} as described above and then determining {the mass range each particle represents} by subtracting the cumulative mass {of the} {particle with the next smallest mass from the cumulative mass of the current particle.  Each particle was then given a weight in the model equal this the amount of cumulative mass space the particle represents.}

\subsection{Violent Creation Model}
{While the ``Basic'' model described in the previous section provides a basic framework to simulate a catastrophic event, we also investigate the effect of the speed distribution of source particles to provide a more realistic scenario for a ``Violent Creation'' of the Geminids.} To model a this violent creation event, the particles were released {with a speed distribution that has a maximum speed {on the order} of 1 km/s} relative to the parent body.  {Most of the implementation details for this Violent Creation Model are the same as for the {B}asic {M}odel with the exception of the additional particle velocity and  that the 10,000 simulated particles were now only split among 100} $\beta${-values instead of each particle having a unique }$\beta${-value}.

{To simulate a more violent creation mechanism, we initialize particles with an isotropic angular distribution with respect to their parent body. We use a statistical sampling process to create this isotropic velocity distribution where we determine the} direction of the velocity vector {for each individual particle} by sampling points randomly within a unit cube and rejecting points that lied outside the unit sphere. {After this procedure, we have a set of isotropically distributed unit vectors representing the additional velocity boost particles gained by the violent creation event. We then determine the absolute speeds of each particle following \citet{durda:96a}, described below. This provides an experimentally-driven model for the expected velocity distribution. We note that it is not possible to directly replicate asteroidal breakup conditions in the laboratory setting and must assume the relation used here scales to our specifically simulated regime}.  The {``typical" }ejection speed $V_a$ {for a particle with mass} $m$ was determined using the power law:
\begin{equation}
    V_a = Cm^{-r}
\end{equation}
where ${C= 0.182}$ {m/s/g}$^{0.1}$ was picked such that the {average (un-weighted)} typical speed {of the simulated particles} was scaled to $\sim$ 1 km/s {and} {where} ${r=0.1}$.  A random scale factor, $x$ was then chosen uniformly between 0 and 4.  A second random number, $y$ uniformly chosen between 0 and 1 was then compared with the Maxwellian function:
\begin{equation}
    (x^2e^{-x^2+1})^w
\end{equation}
where $w$ was picked to equal 1.  If that second random number, $y$ was less than the Maxwellian function, then the particle is assigned the velocity $V=xV_a$, otherwise $x$ was rechosen.

Unlike in the Basic Model, only 100 $\beta$-values were evenly chosen between $\beta = 0$ and $\beta = .052$.  At each of these $\beta$-values, 100 particles were simulated, each with separately chosen velocities.

\subsection{Cometary Model}
For the {C}ometary {M}odel, particles were released at locations throughout a single orbit of (3200) Phaethon {2000 years ago}.  Due to the increased complexity of this simulation and the frequency at which particles escaped or spiraled into the Sun, the number of particles simulated was increased to 100,000 for this model.  Particles were released at 100 locations, 50 on either side of perihelion, equally spaced in their radial distance from the Sun.  At each location, the maximum theoretical $\beta$-value for bound particles was determined and 1000 particles were released at a relative velocity of 0 m/s relative to (3200) Phaethon, evenly spaced in $\beta$-space.  {The model weight,} $w$, {is assigned to each particle as follows:}
\begin{equation}
    {w = \frac{Mt}{r^4}}
\end{equation}
{where} $M$ {is the amount of cumulative mass space represented by the particle, similar to the weighting functions of the Basic Model and Violent Creation Model,} $r$ {is the distance between the particle location and the Sun and is to the} $-4$ {power following \citet{ryabova:07a}, and} $t$ {represents the amount of time (3200) Phaethon spent closest to this release location as opposed to one of the other 99 modeled release locations.}



\section{Results} \label{comparison}

\subsection{Stream Characteristics}
\begin{figure}
    \centering
    \includegraphics[width = .95\textwidth]{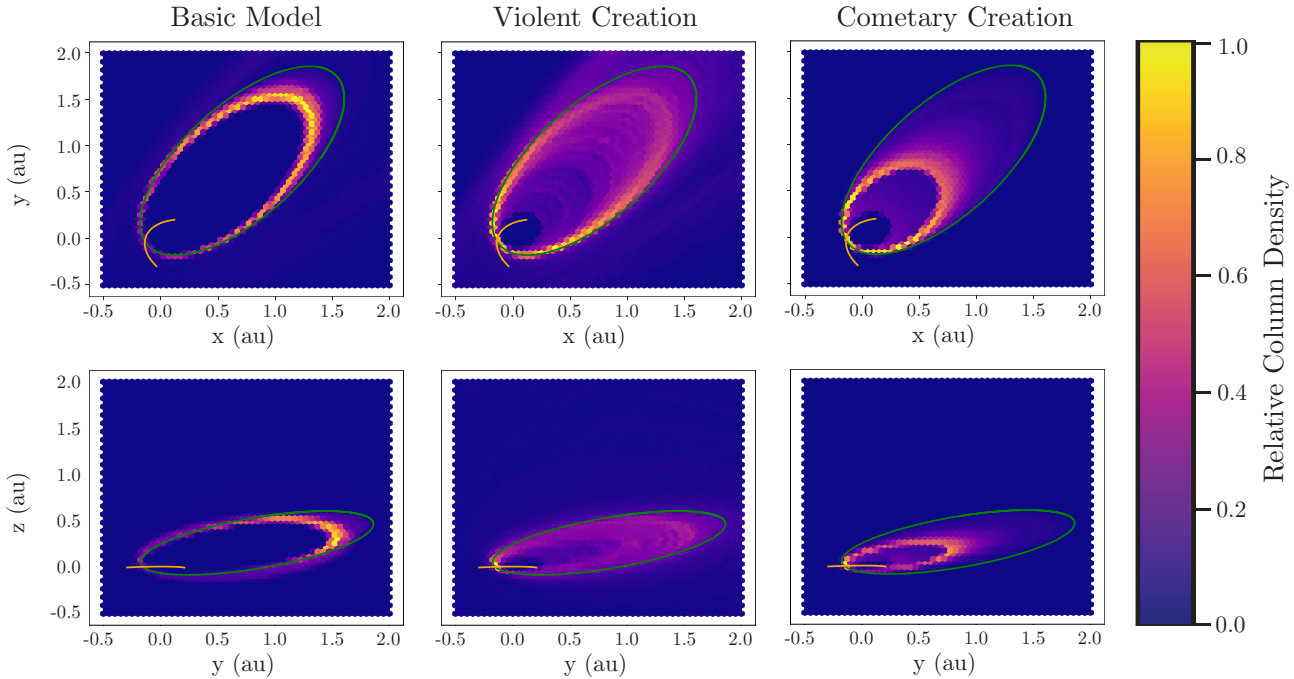}
    \caption{{Number c}olumn density of the Geminids stream in the x-y (top) and y-z (bottom) planes as described by the basic model (left) and Violent Creation Model (mid), and Cometary Creation Models (right) {during the last two years of simulation.}  Densities are normalized to 1 in each panel independently.  The green ellipse is the orbit of (3200) Phaethon, the orange curve is the orbit of PSP during the anomalous spike in dust flux.  All panels are in Ecliptic J2000 coordinates.  Note that particles with a semi-major axis of $a<0.2$ were removed from the simulations leading to the discontinuity in density near the {S}un in the {C}ometary {C}reation {M}odel.}
    \label{fig:column}
\end{figure}

\begin{figure}
    \centering
    \includegraphics[width = .95\textwidth]{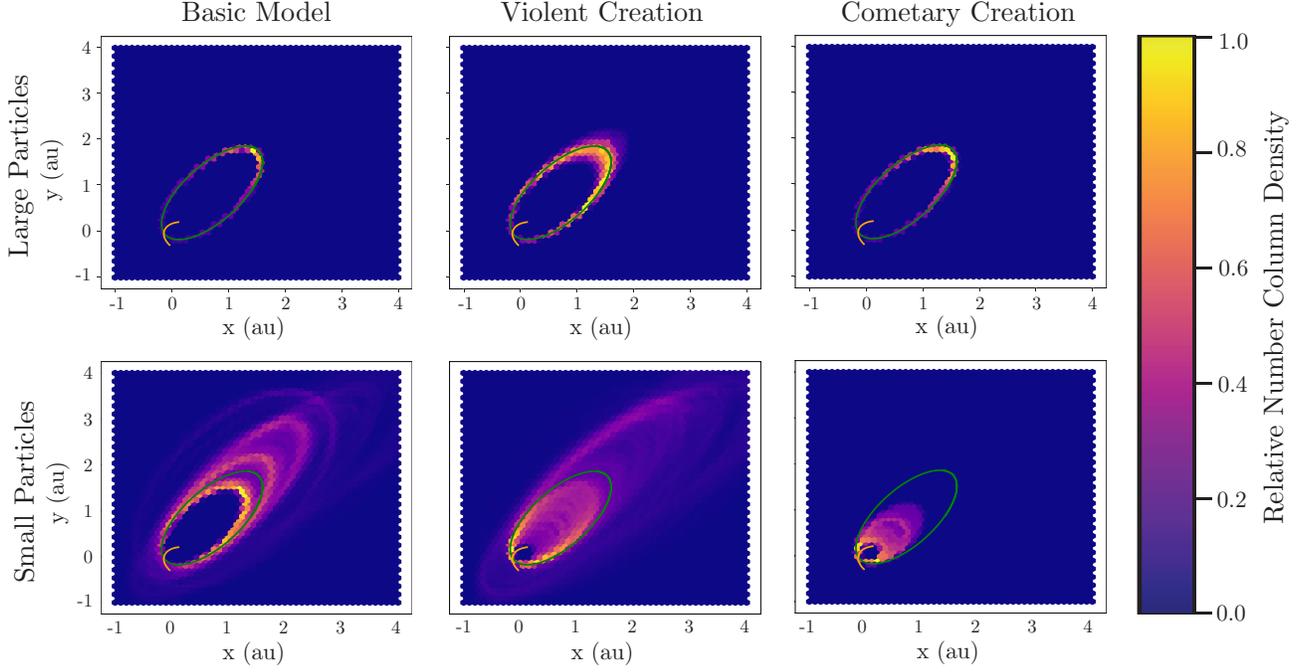}
    \caption{{Number column density of the Geminids stream in the x-y plane as described by the Basic Model (left) and Violent Creation Model (mid), and Cometary Creation Model (right) during the last two years of simulation.  Densities are normalized to 1 in each panel independently.  The green ellipse is the orbit of (3200) Phaethon, the orange curve is the orbit of PSP during {the} anomalous spike in dust flux.  All panels are in Ecliptic J2000 coordinates.  Note that particles with a semi-major axis of} $a<0.2$ {were removed from the simulations leading to the discontinuity in density near the Sun in the Cometary Creation Model.  The top panels show ``large particles'' which are defined as particles with }$m > 10^{-5}$ {g, which are the particles that we consider at Earth in Section \ref{sec:earth}.  The bottom panels show "small particles" which are defined as particles with }$m < 10^{-7}${ g.}}
    \label{fig:size}
\end{figure}

\begin{figure}
    \centering
    \includegraphics[width = \textwidth]{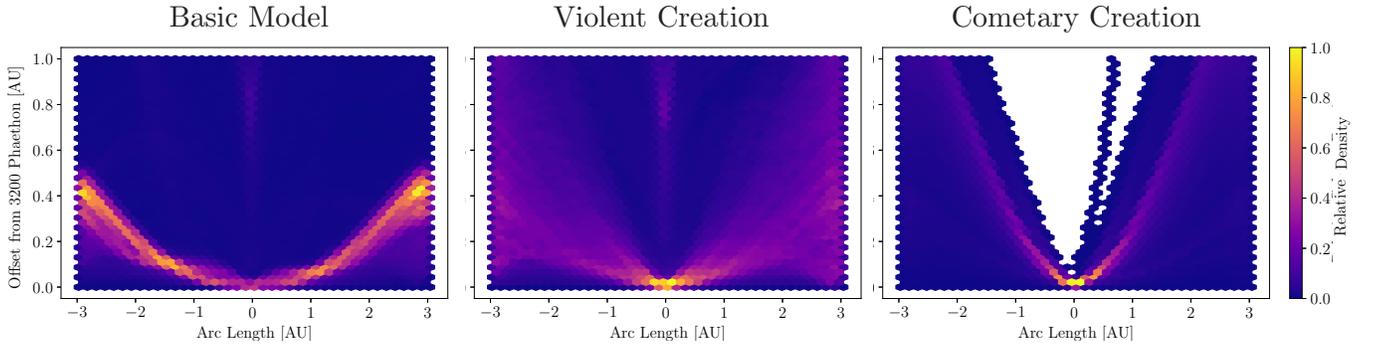}
    \caption{The relative number of particles located at each point around the "unwrapped" orbit of (3200) Phaethon for the Basic Model (left) Violent Creation Model (middle) and the Cometary Creation Model (right) {during the last two years of simulation.}  To create this figure each particle was radially projected onto the orbit of (3200) Phaethon and then the distance between that point and the particle was calculated.  {The white area in the rightmost panel is locations where no particles exist and is caused by some combination of the removal of particles that are too close to the sun and the fact that little to no particles exist outside the orbit of (3200) Phaethon in the Cometary Model.}}
    \label{fig:specto}
\end{figure}

\begin{figure}
    \centering
    \includegraphics[width=0.8\textwidth]{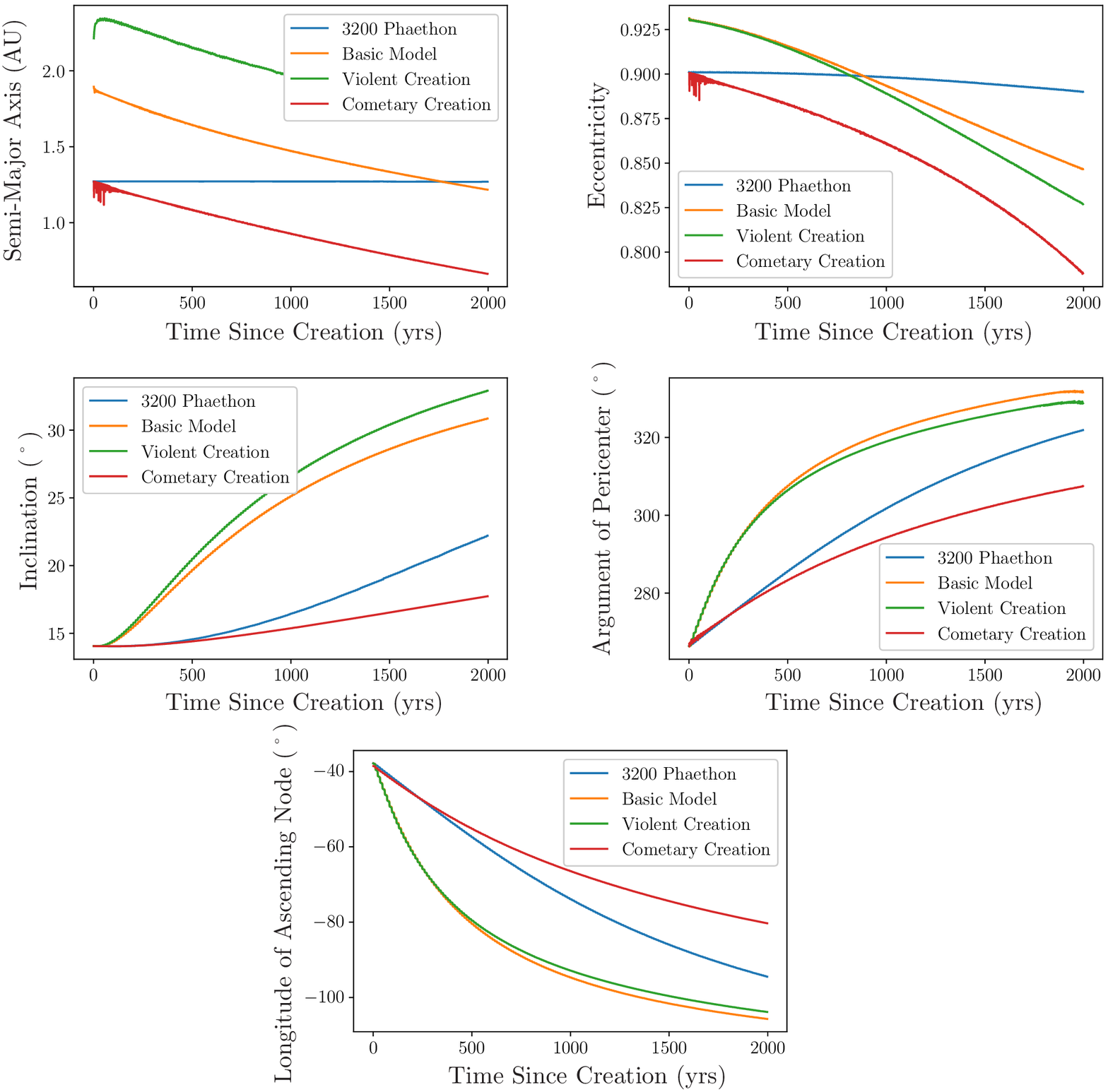}
    \caption{Time evolution of the orbital elements of the average particle in the {primary stream} ($m>10^{-8}$ {g) from the} model variants and for (3200) Phaethon.  All elements are measured in the Ecliptic J2000 frame.  Top Left: Semi-Major axis. Top Right: Eccentricity. Middle Left: Inclination ($^\circ$).  Middle Right: Argument of Pericenter ($^\circ$). Bottom: Longitude of Ascending Node ($^\circ$).}
    \label{fig:elements}
\end{figure}


Column densities of the modeled streams in the Ecliptic J2000 frame {during the last two years of simulation} are presented in Figure \ref{fig:column}. Note these show column density, not 3D density, and thus represent the number of particles per area along a particular line of sight.  {To demonstrate how the particle size varies throughout the stream we additionally plot column densities for extremely small particles} ($m<10^{-7}$ g) {and large particles} ($m > 10^{-5}$ g) {in Figure \ref{fig:size}.}  We also present plots of the ``unwrapped'' stream in Figure \ref{fig:specto}.  These quantities were generated by projecting the particles to the orbit of (3200) Phaethon and then calculating the distance from the particle to that point against the arc length distance of that point from perihelion.

The {B}asic {M}odel has a relatively dense central core to the inside of the the orbit of (3200) Phaethon and a less dense outer stream composed of lighter particles that is outside the orbit of (3200) Phaethon.  As expected, particles that orbit near the orbit of (3200) Phaethon {are} generally the most massive and the mass of particles decreases further {inwards} from the orbit of the parent body.  There additionally appears to be a secondary stream outside of the orbit of (3200) Phaethon composed of {extremely} small particles at a larger radius.  Particles with $\beta < 0.02$ generally were dominated by Poynting-Robertson drag and had a final orbit inside that of (3200) Phaethon while those particles with $\beta>0.02$ were boosted to large enough orbits by radiation pressure that after the 2000 year simulation they still had orbits larger than that of (3200) Phaethon, leading to the development of a secondary stream.  {The fact that this secondary stream is composed primarily of light particles, however, means that this stream only accounts for about }$0.07\%$ {of the total mass of the stream.}

The Violent Creation Model yields a more dispersed stream than the Basic Model due to the initial additional velocity spread for the grains.  {The densest part of the stream occurs near perihelion as most particle spread happens near aphelion. As seen in Figure} \ref{fig:size}, this model leads to a {relatively} more homogeneous distribution of particle size throughout the stream {compared to the Basic Model}, likely due to the mixing caused by the initial speeds given to each particle at the start of the simulation.  The average final semi-major axis of a particle at a given value for $\beta$ follows the same trend as for the Basic Model but the variation within each $\beta$-bin is far greater than the variation between $\beta$-bins, especially for $\beta < 0.02$.

The Cometary Creation Model yields a stream in which most of the particles orbit inside the orbit of (3200) Phaethon.  This is likely due to the fact particles released at further distances from the Sun will experience a smaller boost in initial semi-major axis due to radiation pressure so the effects of Poynting-Robertson Drag will more easily dominate.  T{his effect is most pronounced for small particles which have larger Poynting-Robertson Drag and thus shorter decay timescales.}  When running the simulations of this model, we noticed that {before weighting,} $\sim 85\%$ of the surviving particles after 2000 years had a $\beta < 0.05$.  There was a small cluster of remaining particles at $\beta=0.3$ and the rest were about evenly distributed over the $\beta$-range.  To increase the resolution of the model and to focus more on detectable particles, we reran the simulation with a maximum $\beta$-value of $0.05$, {corresponding with a minimum mass of }$2.8 \times 10^{-9}$ {g.  We use the results of this} $\beta<0.05$ {limited simulation instead of the full} $\beta${-range in all of our analysis and figures.}

 The key difference between the models is the Basic Model expects a well-defined stream just inside the orbit of (3200) Phaethon, the Violent Creation Model expects a more dispersed stream but still centered on the same orbit as the Basic Model, and the Cometary Creation model predicts a stream that has decayed significantly and, while spread out, has a {stream core, which we define as the region with the greatest particle density}, with a semi-major axis $a<1$ AU.

We additionally present the evolution of orbital elements of the stream over time in Figure \ref{fig:elements}.  These were made by taking the average of the orbital elements for each surviving particle {with mass} $m> 1 \times 10^{-8}$ {g, weighted by the model weights described above, for} each year during the simulation.  T{his limiting mass is applied to limit our analysis to the primary stream. } Clearly demonstrated in these plots is the {effect of Poynting-Robertson drag} circularizing the orbits of the streams {as seen by} the average semi-major axis and eccentricity for all three models being driven down over time.  Note that for the {C}ometary {C}reation {M}odel in particular, the change in eccentricity appears to be accelerating.  The differing initial semi-major axis and eccentricities for the {B}asic {Model} and {V}iolent {C}reation {M}odel is caused by the radiative forces being suddenly turned on at perihelion, as is done in our models, causing an immediate shift the aphelion positions of the particles.  This effect is not seen in the {C}ometary {C}reation {M}odel because the instantaneous effect of radiative forces is not as significant for particles released further away from the Sun.  We also note that the {C}ometary {C}reation {M}odel exhibits slower precession than (3200) Phaethon while the {B}asic {Model} and {V}iolent {C}reation {M}odels exhibit faster precession than their parent body.  {An additional effect likely demonstrated by the model is the Kozai-Lidov mechanism where the reduction in eccentricity leads to an increase in the inclination of the model \citep{1962AJ.....67..591K}.}
\subsection{Geminids Stream at Earth} \label{sec:earth}
\begin{figure}
    \centering
    \includegraphics[width = .5\textwidth]{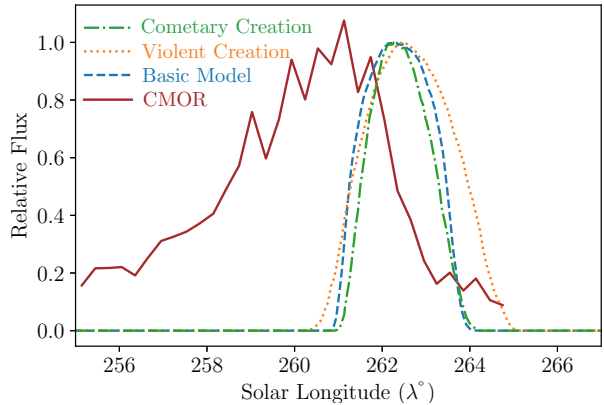}
    \caption{Plot of the normalized impactor flux upon Earth as predicted during the last two years of simulation by the {Basic Model (blue dashed line), the Violent Creation Model (orange dotted line) and Cometary Creation Model (green dot-dashed line) compared to observational data taken from CMOR (solid red line) \citep{jones:05a}.}}
    \label{fig:flux}
\end{figure}

\begin{table}
    \centering
    \begin{tabular}{l|cccc}
        Model &  Mass (g) & Peak Time ($\lambda^\circ$) & {Full-Width, Half-Max} ($\lambda^\circ$) & Closest Approach (au) \\
        \hline
         Basic & ${4.7 (\pm 3.7)\times 10^{14}}$  & 262.26 & {2.34} & 0.019\\
         Violent & ${6.3 (\pm 3.7)\times 10^{13}}$ &{262.51} & {2.79} &{0.0017}\\
         Cometary & ${8.6 (\pm 4.3)\times 10^{14}}$ &{262.14} & {1.97} & 0.017\\
    \end{tabular}
    \caption{Predicted and derived physical characteristics of the stream at Earth for each of the models.  The peak flux observed by CMOR occurs at $\sim 261.1 ^\circ$.  {The error in Geminids mass was calculated by propogating the error in CMOR flux and the Geminids power law index taken from \citet{blaauw:17a}}.}
    \label{tab:earth}.  
\end{table}
To normalize the model results to real physical units, we compare with Earth-based observations of the Geminids stream.  To do so, we calculate the density of meteoroids  in a 0.03 AU sphere centered on Earth at each {of 8000} points {along} its orbit.  {Note that the selection of sphere radius does add uncertainty to the stream mass and timing and can substantially alter the full-width, half-maximum of the stream.  We chose a .03 AU sphere as the Basic Model and Cometary Creation Model do not register more than a couple particles for a .02 AU sphere.}  We then calculate the flux incident on Earth by treating Earth as a flat plate detector perpendicular to the stream of meteoroids that are incident upon it at $35$ km/s.  The normalized flux is compared to that of CMOR as shown in Figure \ref{fig:flux}.  The minimal limiting mass for the meteoroids is taken to be $1 \times 10^{-5}$ g whereas the limiting mass for CMOR is $1.8 \times 10^{-4}$ g following \citet{blaauw:17a}.  The lower mass limit allows the {C}ometary {Creation} {M}odel to be detected at Earth as no particles from the {C}ometary {Creation} {M}odel with mass above $1.8 \times 10^{-4}$ g were detected within 0.05 AU of Earth.  This lower limiting mass will cause the stream mass to be underestimated and the peak time to be slightly earlier. {We note there are multiple other observations of the Geminids at Earth \citep[e.g.][]{kero:13a} and the Moon \citep[e.g.][]{Szalay:18a}, however, we limit the comparison to CMOR as a nominal example of the profile and as these measurements were used to derive the total stream mass.}

The stream mass for each model is estimated by scaling up the peak meteroid flux observed at Earth to the peak flux observed by CMOR as follows:
\begin{equation}
{
    M_{\text{geminids}} = M_{\text{model}} * \frac{F_{\text{CMOR}}}{F_{\text{model}}}*\frac{ m_{\text{model limit}}^{1-s}-m_{\text{max}}^{1-s}}{ m_{\text{CMOR limit}}^{1-s}}}
\end{equation}
{where} $M_{\text{model}}$ {is the weighted sum of the mass of the 10,000 particles used in the model,} $F_{\text{model}}$ {is the weighted particle flux upon Earth as predicted by the model, }$F_{\text{CMOR}} = {6.64} \times 10^{-2}$ meteors /km${}^2$/hr-- {which was determined by} scaling the {off-peak flux} given in \citet{blaauw:17a} to the {time of} peak flux, ${m_{\text{max}} = 10}$ {g is the maximum mass considered by the model,} $m_{\text{CMOR limit}}$ {and} $m_{\text{model limit}}$ {are the limit{ing} values for CMOR and the model respectivly as discussed above, and} $s=1.68$ i{s the mass power law we have used throughout this paper.  The last fractional part of this equation is derived from Equation} \ref{eqn:cume_mass} {and used to} account for the fact that our model theoretically registers a greater fraction of the {G}eminids meteoroids than CMOR due to the lower limiting mass.  {The masses estimated by this model} range from ${6.3 \times 10^{13}}$ g for the {V}iolent {C}reation {M}odel to ${8.6} \times 10^{14}$ g for the {C}ometary {Creation M}odel.  These values are about one half to one order of magnitude smaller than the other estimates of the mass of the Geminids stream when adjusted for the limiting mass ranges of {previous} estimates \citep{ryabova:17a,blaauw:17a}.  A comparison of each model at Earth with observational CMOR data is visible in Figure \ref{fig:flux}.  A summary of physical characteristics of the stream at Earth is visible in Table \ref{tab:earth}.

The flux at Earth for each of these models peaks at approximately $262.2^\circ$ for each of the models.  While this is in agreement with the traditional visual peak of the Geminids at $262.2^\circ$, it disagrees with CMOR observations by about $0.9-1.3$ days dependent on model as CMOR peaks at about $261.4^\circ$.  {CMOR observes meteors with magnitudes down to $+8$, which are fainter than the more massive and therefore brighter and human-eye visible meteors that peak later.}  Given the models' limiting masses are designed to mimic CMOR's detection limit, the likeliest explanation is model uncertainty with respect to precise formation mechanism.

A key thing to note is that none of the models were able to have any particles collide with Earth--the Violent Creation Model had{ a }closest approach distance of .0024 AU while the Basic {Model} and Cometary Creation {M}odel had  closest approaches on the order of 0.02 AU.   Given that only 10,000 or 100,000 particles were simulated dependent on model, this fact is likely in part due to low sample sizes, particularly for the {V}iolent {C}reation {M}odel.  We also only consider 1000 points in space for each particle over a two year orbit which impacts the spacial resolution data.   Notably, the meteoroids from the {V}iolent  {C}reation {M}odel had a closest approach to Earth an order of magnitude closer than the other models, likely due to the randomness included in the {V}iolent {C}reation {M}odel allowing for more particles to closely approach Earth. This inability to exactly model the imapct at Earth has been exhibited by other models in the literature \citep{ryabova:07a,ryabova:17a} which likely indicates a not yet fully understood or postulated mechanism.

\subsection{Space Missions}
\subsubsection{Parker Solar Probe}
\begin{figure}
    \centering
    \includegraphics[width = 0.5\textwidth]{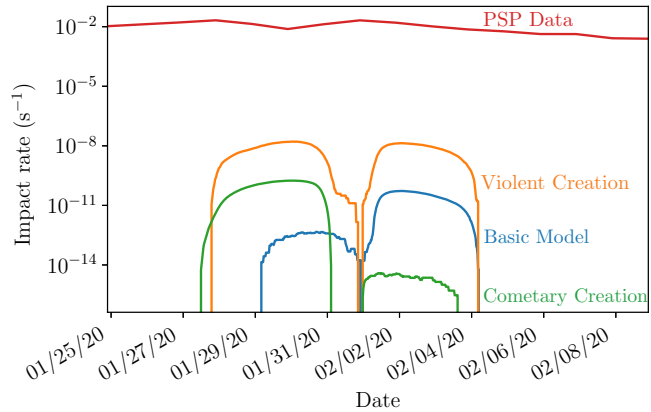}
    \caption{Dust flux incident on the Parker Solar Probe from the Geminids stream around perihelion of orbit 4 of PSP as predicted {during the last two years of simulation} by the various models and compared with the observed dust flux showing the post perihelion enhancement.  Dust flux {was} modeled using the density of meteoroids within 0.05 au of PSP. { PSP impact rates {are} from previously published observations} \citep{szalay:21a}.}
    \label{fig:psp}
\end{figure}
\begin{figure}
    \centering
    \includegraphics[width=0.9\textwidth]{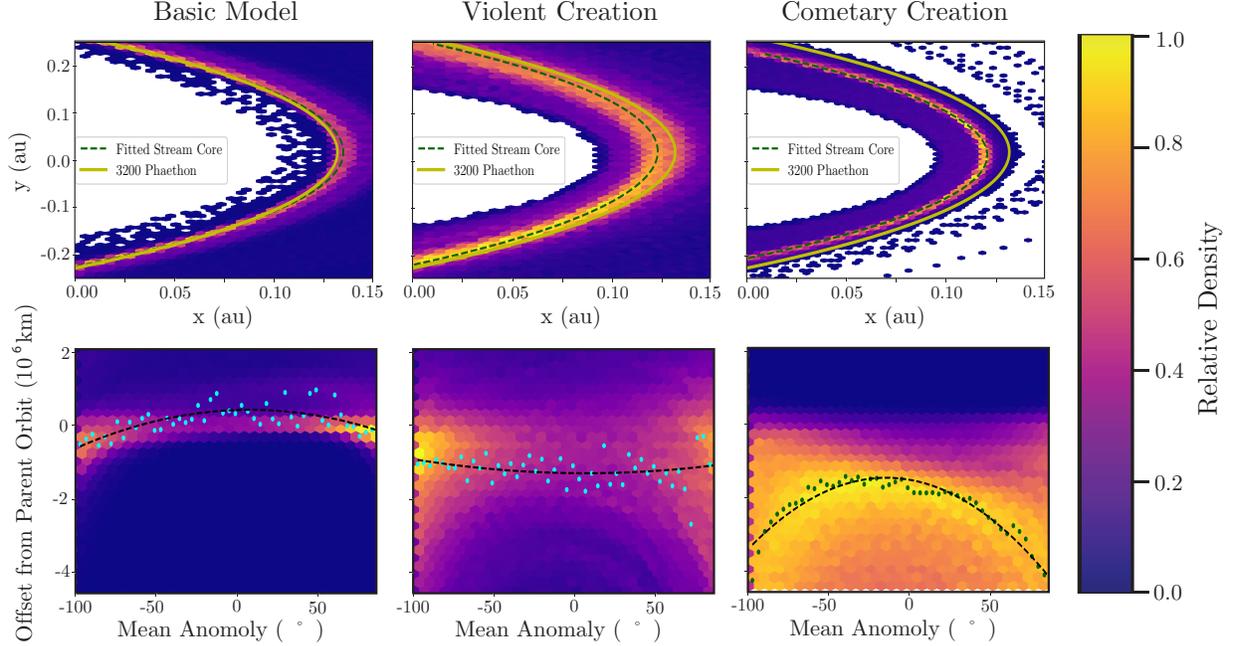}
    \caption{Top: {Number c}olumn Density plots in the plane of the orbit of (3200) Phaethon for the Basic Model (left) Violent Creation Model (mid) and the Cometary Creation Model (right) {during the last two years of simulation}.  Shown over top these plots is the orbit of (3200) Phaethon (yellow solid line) and the fitted core of the modeled stream (dashed green line).  Bottom: The relative number of particles at each signed offset from the orbit of (3200) Phaethon as a function of mean anomaly.  {Cyan and dark green points in the bottom panel} show the center of the radial bins of maximum density for each mean anomaly bin.  The dashed line is a parabolic fit to those points and is taken to represent the core of the stream.  Coefficients of the parabolic fit can be found in Table \ref{tab:coeffs}.}
    \label{fig:psp_peri}
\end{figure}
\begin{table}[]
    \centering
    \begin{tabular}{l|ccc}
    Model & a & b & c \\
    \hline
    Basic &$-91 {\pm 15}$ & $1.4 {(\pm 0.7)} \times 10^{3} $ & $4.3 {(\pm 0.6)}\times 10^{5}$ \\
    Violent Creation & {34} ${\pm 22}$ & ${-400 \pm 1100}$ & ${-1.3 (\pm 0.9) \times 10^{5}}$\\
    Cometary Creation & -270 ${\pm 10}$& $-7.9 {(\pm 0.6)} \times 10^{3}$ & $-1.6 {(\pm 0.5)}\times 10^{6}$\\

    \end{tabular}
    \caption{Coefficients of the fitted parabola $y = ax^2 + bx + c$ where $x$ is the mean anomaly in degrees and $y$ is the signed offset from the orbit of (3200) Phaethon of the core of the meteoroid stream.  These fits are shown in Figure \ref{fig:psp_peri}.}
    \label{tab:coeffs}
\end{table}

Similar to how the flux at Earth was calculated, the flux incident on the Parker Solar Probe was estimated by determining the density of dust particles within 0.05 AU of PSP at {each of 8000 points along PSP's orbit} and then calculating how many particles PSP would hit if it flew through a cloud of particles at that density.  To achieve an upper bound on that calculation we assumed PSP was a flat plate dust detector facing perpendicular to  {every particle and} moving at a speed relative to the particles equal to the sum of the orbital speed of PSP with respect to the Sun and the speed of (3200) Phaethon at perihelion.  The {total mass of the} stream for this calculation was {set} to $10^{14}$ g.  {We additionally assume that every particle, regardless of mass, registers on PSP with 100\% efficiency.} Using these assumptions we find that the impact rate was on the order of $10^{{-8}}$ s$^{-1}$ for the{ V}iolent {C}reation {M}odel and on the order of ${10^{-9}-10^{-11}}$ for the {B}asic {M}odel and {C}ometary {C}reation {M}odel, meaning that our models of the Geminids do not provide sufficient impact rates to explain the post perihelion rate enhancement observed by PSP \citep{szalay:21a}. The predicted impact rate due to the Geminids stream to PSP is shown in Figure \ref{fig:psp}. {This figure also shows in-situ impact rate data from PSP's orbit 4, as {a} representative orbit where the second peak is attributed to a possible Geminids-related source \citep{szalay:21a}}.

We additionally present a more detailed prediction of stream observations near perihelion.  We present in Figure \ref{fig:psp_peri} column densities in the plane of the orbit of (3200) Phaethon along with a fitted ``core" of the stream.  Additionally shown is the plane-projected offset of the stream from Phaethon's orbit as a function of the mean anomaly.  Fitting the core of a point cloud is an imprecise task.  To do so we first binned the particles by mean anomaly.  These bins were then subdivided into bins based on the radial distance, in the plane of the orbit of (3200) Phaethon, from the orbit of (3200) Phaethon.  For each mean anomaly bin, the radial bin with the peak density was chosen to be representative of the stream.  To create a continuous fit and to facilitate comparison with \citet{battams:22a}, we then fit a parabola that took in mean anomaly as the independent variable and returned the radial offset from the orbit of (3200) Phaethon to these representative points.  The coefficients of the fit equation $y = ax^2 + bx + c$ for each model are given in Table \ref{tab:coeffs}.

We note that stream for the most part lies to the inside of the orbit of (3200) Phaethon, with the only exception being the Basic Model very close to perihelion.  Additionally, none of the models are symmetric around perihelion, the streams of the Basic Model and the Violent Creation Model are both to first order further from the Sun  just after perihelion compared to just before perihelion while the Cometary Creation Model is the opposite.  {The Basic Model and Cometary Creation Model}  are concave down to second order meaning that the further from perihelion, the more interior the particles are, {while the Violent Creation Model is instead concave up}. Observations from PSP near the perihelion of (3200) Phaethon as reported by \citet{battams:22a} show that contrary to our {Basic Model and Cometary Model}, the stream is in large part outside the orbit of (3200) Phaethon and concave up to second order.  These observations do, however, agree with our model in the asymmetry about perihelion.  Overall, these observations are most consistent with our Basic Model which is exterior to the orbit of (3200) Phaethon at perihelion and our {V}iolent {C}reation {M}odel which {is concave up around perihelion}.

\subsubsection{DESTINY+}
The JAXA/DESTINY+ mission is planned to rendezvous with (3200) Phaethon \citep{sarli:18a}. Onboard it carries the DESTINY+ Dust Analyzer (DDA), a dust detector capable of compositional determination of impacting dust grains with an effective area of 350 cm$^2$ \citep{kobayashi:18a}. In addition to the direct observations of grains from Phaethon's impact ejecta cloud \citep{szalay:19a}, DESTINY+ will make important observations of interplanetary and interstellar dust during its cruise phase \citep{kruger:19a}. 

We simulated the passage of an orbit similar to DESTINY+ through the Geminids stream and found that although the rate of impacts might be detectable over the background level, the total expected detections throughout the mission of Geminids grains was less than 1 impact. Hence, we anticipate DESTINY+ will likely be moving too rapidly through the core of the stream to have any Geminids dust hit its dust analyzer.

\section{Discussion and Conclusions} \label{conclusions}
We have simulated the Geminids meteoroid stream using a number of models and analyzed the Geminids flux upon Earth, Parker Solar Probe, and DESTINY+. Comparing to remote imaging of the Geminids stream \citep{battams:22a}, we find the observed stream location external to the orbit of (3200) Phaethon is most consistent with a formation mechanism of a rapid low-speed release of material near perihelion, not by a more temporally extended cometary formation mechanism. Hence, this comparison suggests the Geminids may have formed via a more violent, catastrophic destruction of bodies that transit very near to the Sun \citep{granvik:16a,maclennan:21a}. {This is also consistent with the cross-comparison of mass of the Geminids steam and mass of (3200) Phaethon \citep{blaauw:17a,ryabova:17a}, which are comparable and suggestive of a catastrophic origin. Similarly, the weaker Daytime Sextantids meteoroid stream, suggested to be part of the Geminid-Phaethon stream complex, has been estimated to also be comparably massive to its proposed parent body asteroid 2005 UD \citep{kipreos:22a}, also consistent with a catastrophic common origin for both  {streams}.} 

{Our models, however,} fail t{o directly hit Earth or} to replicate the exact timing {of the Geminids shower}-- the peak flux at Earth is just less than 1.5 days after the observed peak for the models.  This discrepancy which has been reported in other models of the Geminids stream needs more investigation to determine its source.  We additionally report mass estimates in the range of ${6.3 \times 10^{13}}$ g to ${8.6 \times 10^{14}}$ g which are {respectively slightly lower than and consistent} with the low side other estimates in the literature \citep{ryabova:17a, blaauw:17a}. 

It is important to note that our model {assumed} particles with an ``asteroidal'' composition defined by \citet{wilck:96a} {and bulk density} {$\rho = 3205$ kg/m$^3$}.  Changes in the composition of the particles can change the $\beta$-values and thus drastically affect the weighting functions.  Additionally, the prediction of meteoroid impacts was done by taking the density of a sphere on the order of a few hundredths of an AU, which is significantly larger than any of the detectors considered.  The mass, timing, and duration of the stream at {Earth} are all sensitive to the choice of the radius of this sphere--we have tried to pick a value which is reasonable and still captures information about the stream locally while not losing too much data to small number statistics.  

We find the Geminids stream is likely not observable {via in-situ impact detections} with the Parker Solar Probe, at least with realistic upper limits on the mass of the Geminids stream.  We additionally have shown that the formation mechanism of the stream dramatically changes the final shape and width of the stream.  By {further} observing stream morphology {via direct imaging to resolve the spatial structure} with PSP near {perihelion} or {with} other future missions, we likely would be able to significantly constrain formation mechanisms.
We also have demonstrated some properties of the Geminids stream such as how Poynting-Robertson drag has led to a reduction in the semi-major axis of particles and segregates them based on size--lighter particles generally exist on the inside edge of the stream while heavier particles mostly exist of the outside edge of the stream with the extremely light particles creating a secondary stream exterior to the orbit of (3200) Phaethon.

An alternative explanation for the PSP post-perihelion impact rate enhancement is that it is the result of the collisional byproducts, a $\beta$-stream, from the Geminids as that was not modeled here and the proximity of PSP to the Geminids makes it still a possible candidate \citep{szalay:21a}.  Future work can use images from the WISPR instrument on PSP \citep[e.g.][]{battams:22a} to compare with and refine our models of the Geminids stream.  More modeling could be done to understand the dust impacts upon PSP, such as understanding if it would be possible to directly observe a meteoroid stream, or the Geminids stream in particular, by flying through its core. Additionally, as the DESTINY+ mission is visiting (3200) Phaethon, data from that mission will likely help us understand the Geminids stream and other meteoroid streams in general. 

Future observations that are able to measure asymmetries in the stream before and after perihelion and which compare the position of the stream to that of the orbit of (3200) Phaethon can likely constrain the formation mechanism further.  Comparisons between the direct visual imaging of the Geminids by PSP \citep[e.g.][]{battams:20a,battams:22a}, notably the location and the width of the stream at multiple locations, could be used to further constrain the origin and evolution of the Geminids, utilizing the stream as a valuable laboratory to investigate meteoroid stream formation and dynamics throughout the solar system.



\section*{Software Availability}
All code used for this paper is available at \url{https://github.com/wcukier/Phaethon_Meteoroids}
\section*{Acknowledgements}
We thank Petr Pokorn\'y for helpful modeling discussions and Harald Kr\"uger for discussions on the DESTINY+ trajectory. We acknowledge the Parker Solar Probe Guest Investigator Program, grant 80NSSC21K1764. WC would like to thank the Hewlett Foundation Fund and the Princeton Undergraduate Fund for Academic Conferences for funding relevant conference travel. 

\software{ 
SpiceyPy \citep{Annex2020SpiceyPyPythonicWrapper},
SPICE \citep{Acton2018LookFutureHandling, Acton1996AncillaryDataServices},
Rebound \citep{rebound},
Reboundx \citep{2020MNRAS.491.2885T},
Matplotlib \citep{Hunter:2007},
Numpy \citep{harris2020array},
Scipy \citep{2020SciPy-NMeth}
}





\bibliographystyle{aasjournal}



\end{document}